\documentclass{article}
\usepackage{graphics}
\usepackage{epsfig}
\usepackage{epstopdf}
\usepackage{amsmath}
\usepackage{float}
\usepackage{caption}
\usepackage{subcaption}
\usepackage{multirow}
\begin{document}
\begin{center}
	{\Large{ Bayesian analysis of running holographic Ricci dark energy}} \\[0.2in]
	Paxy George and Titus K Mathew \\
	Department of Physics, \\
	Cochin University of Science and Technology, Kochi-22, India.
\end{center}
\begin{abstract}
	 Holographic Ricci dark energy evolving through its interaction with dark matter is a natural choice for the running vacuum energy model.
	We have analyzed the relative significance of two versions of this model in the light of SNIa, CMB, BAO and Hubble data sets
	using the method Bayesian inferences. The first one, model 1, is the running holographic Ricci dark energy (rhrde) having a constant
	additive term in its density form and the second is one, model 2, having no additive constant, instead the interaction of rhrde with
	dark matter is accounted through a phenomenological coupling term. The Bayes factor of these models in comparison with the
	standard $\Lambda$CDM have been obtained by calculating the likelihood of each model for four different data combinations,
	SNIa(307)+CMB+BAO, SNIa(307)+CMB+BAO+Hubble data, SNIa(580)+CMB+BAO and SNIa(580)+CMB+BAO+Hubble
	data. Suitable flat priors for the model parameters has been assumed for calculating the likelihood in both cases.
	Our analysis  shows that, according to the Jeffreys scale,  the evidence for $\Lambda$CDM against both model 1 and model 2 is very strong as the Bayes factor of both models are much less than one for all the data combinations. 
\end{abstract}
\section{Introduction}

The idea that the cosmological term $\Lambda$ could be varying (at least slowly) as the universe expands is gaining much attention
in the light of the recent cosmological data \cite{doi:10.1142/S0217751X16300350}. A continuously varying vacuum energy which
adopts its time dependence from the Hubble parameter $H(t)$ is suggested by quantum field theory in curved space-time
\cite{2006MPLA...21..479S} and it turns out to be a good choice for dynamical cosmological term
\cite{Sol__2015,Sol__2013,doi:10.1142/S0218271815410035,Sol__2017}. Such a running vacuum energy (RVE), with constant equation 
of state, $\omega=-1,$ model has been proposed
to alleviate the drawbacks of the standard $\Lambda$CDM \cite{Copeland:2006wr,Bahcall:1999xn},  the coincidence problem and the
cosmological constant problem. It was proposed to explain the recent acceleration in the expansion of the universe
\cite{Riess:1998cb,1999ApJ...517..565P,Spergel:2003cb,Tegmark:2003uf}.
The issue of the  cosmological constant problem is due to the incredibly small magnitude of the observed value of the cosmological
constant
compared to the theoretical prediction from quantum field theory \cite{amendola_tsujikawa_2010}. 
The coincidence of
present energy densities of both cosmological constant and dark matter 
is referred  to as the coincidence problem.
RVE model \cite{Shapiro:1999zt,Shapiro:2003kv} proposes a cosmological parameter, $\Lambda(H),$ which is decaying as the
universe expands and can hopefully solve the previous mentioned issues. In a recent work \cite{Sol__2017} the authors have found
strong evidence for a slowly varying `cosmological constant' using the latest combined observational data,
SNIa+BAO+H(z)+LSS+BBN+CMB.

Another interesting attempt to explain the decaying dark energy is by applying holographic principle
\cite{RevModPhys.74.825,tHooft:1993dmi,Susskind:1994vu} which has culminated into an alternative model, the holographic
Ricci dark energy (hrde). The hrde has a resemblance in form with the conventional RVE. Energy densities of both the models
are combinations of $\dot{H}$ and $H^2,$ where over-dot represents a derivative with respect to time. The hrde was initially
proposed as a varying dark energy with time dependent equation of state \cite{doi:10.1142/S0218271814500242}. Due to its
similarity in density form, it has been considered as an alternative to RVE
\cite{PhysRevD.87.023515,doi:10.1142/S0217732316500759,George:2018myt} and  for convenience let us name it as
rhrde (running holographic Ricci dark energy).  In the conventional rhrde model
there exist an additive constant in the energy density and is essential to ensure the transition from a prior decelerated to a
late accelerated epoch \cite{PhysRevD.87.023515,doi:10.1142/S0217732316500759}.
We \cite{George:2018myt}  have shown that such an additive constant in the density is not needed for causing a transition
into the late accelerating epoch if one accounts for the interaction between the rhrde and the dark matter \cite{PhysRevD.79.043517,Som:2014hja,He:2008tn}.
The evolution of the various cosmological parameters in such a model were studied in the light of supernovae, SDSS and recent Planck data
\cite{George:2018myt}. Taking account of the reasonable performance of rhrde in predicting the back ground evolution of the
universe, it is worth to contrast it with the standard $\Lambda$CDM model in order to assess its relative importance. In the
present work we make this comparison
using the method of Bayesian analysis \cite{doi:10.1080/01621459.1995.10476572,PhysRevLett.119.101301,SantosdaCosta:2017ctv,PhysRevD.97.083518}.

The Bayesian theory of statistics was developed by great mathematicians like Gauss, Bayes, Laplace, Bernoulli etc.
Usually it is possible to assign a probability for a random variable based on repeated measurements. But in cosmological scenario
such kind of repeated observations are practically impossible or only rarely possible.
Often one proposes a hypothesis or a
theory in cosmology instead of a random variable, the probability of which is to be fixed. Bayesian theory help us in this
regard to fix the probability of such hypothesis by using the available data
\cite{Kurek_2008,Cid:2018ugy,Guimar_es_2009,10.1093/mnras/stt1954,Colistete:2003xx}. Following this, it is possible to compare different
cosmological
models by computing what is known as Bayes factor which is proportional to the ratio of the probabilities of the models
which are to be compared.  This method have been used in many works for model comparison \cite{PhysRevD.95.123514,Jesus_2017,Serra:2007id,PhysRevD.65.043506,Efstathiou:2008ed}.
Here we first compare the rhrde having an additive
constant in the density with the  standard $\Lambda$CDM model and then compare the rhrde devoid of the additive constant,
but its interaction with cold dark matter taken into accounted phenomenologically.

The rest of the article is arranged in the following way. In section 2 we describe the method of Bayesian analysis. In section 3, 
we present a brief discussion about the rhrde models, model 1, the rhrde with an additive constant in its density and model 2 does 
not have such an additive constant, but the interaction between rhrde and dark matter is accounted through a phenomenological term. 
We also present the calculation of the important model parameters. In section 4 and 5 we obtain the Bayes factors of both the models 
by assuming suitable priors for all the parameters in the models. The conclusions are presents in the last section.

\section{The method of Bayesian analysis}
In this section we present the basic method of the Bayesian comparison following reference \cite{PhysRevD.65.043506,John:2005bz}.
The method is based on the famous Bayes theorem, which
allows one to evaluate the posterior probability of a given model. Let $p(M_{\text{rhrde}}|D,I)$ be the posterior probability
of $M_{\text{rhrde}},$  the rhrde model with Ricci dark energy and dark matter as the cosmic components, such that $D,$ the given
set of observational data and  $I,$ the back ground information regarding the expansion of the universe are true\cite{2010MNRAS.408..865T}. According to
Bayes theorem it then follows, 
\begin{equation}\label{equ:post1}
p(M_{\text{rhrde}}|D,I)=\frac{p(M_{\text{rhrde}}|I)p(D|M_{\text{rhrde}},I)}{p(D|I)}.
\end{equation}
Here $p(M_{\text{rhrde}}|I)$ is the prior probability assigned to the model before analyzing the data, given that the back ground
information $I$ is true.
The prior is a fundamental ingredient of Bayesian statistics. It could be considered as
problematic since the theory does not give prescription about how the prior should be selected. It may be natural that different
scientists might have different priors as a result of their past experiences. But once a prior has been chosen
then repeated
application of Bayes theorem, equation~(\ref{equ:post1}) will lead to a convergence to a common posterior.
The term $p(D|M_{\text{rhrde}},I)$ is the probability for obtaining the data provided the model and the background information
are true and is called the likelihood of the model, $\mathcal{L}(M_{\text{rhrde}})$ for a given set of observational data. This term
encodes how the degree of plausibility of the hypothesis or model changes when we acquire new data.
The term $p(D|I)$ is the probability for obtaining the data if the background information is true and is a normalization factor.
In a similar way the posterior probability of the standard $\Lambda$CDM model ($ M_{\Lambda \text{CDM}}$) can be obtained by
replacing $M_{\text{rhrde}} $ with $M_{\Lambda \text{CDM}}$ in the above equation.
For comparing the models in the Bayesian way we define what is called as the odds
ratio, the ratio of the posterior probabilities of the two models as,
\begin{equation}\label{equ:odds}
O_\text{ij}\equiv \frac{p(M_{\text{rhrde}}|D,I)}{p(M_{\Lambda \text{CDM}}|D,I)}=\frac{p(M_{\text{rhrde}}|I)p(D|M_{\text{rhrde}},I)}{p(M_{\Lambda \text{CDM}}|I)p(D|M_{\Lambda \text{CDM}},I)},
\end{equation}
where suffix $'\text{i}'$ represents $M_{\text{rhrde}}$ and $'\text{j}'$ represents $M_{\Lambda \text{CDM}}$ and we have assumed that the
normalization factor $P(D|I)$ is the same in the case of both models.
The prior probability of either the given model or the standard model is fixed before considering the data and it depends only
on the prior information.
The prior information is an attempt to quantify the collective wisdom of a researcher in the given area. If this collective
information does not prefer one model over the other, the prior probabilities get canceled out in the above odds ratio and
hence we have
\begin{equation}\label{equ:odds1}
O_\text{ij}=\frac{p(D|M_{\text{rhrde}},I)}{p(D|M_{\Lambda \text{CDM}},I)}\equiv B_\text{ij},
\end{equation}
where $B_\text{ij}$ is now called  the Bayes factor.
The $p(D|M_{\text{rhrde}},I),$ the likelihood of the model $M_{\text{rhrde}}$ can be conveniently denoted as $\mathcal{L}(M_\text{i})$
(where $M_\text{i}=M_{\text{rhrde}}$ ).
If $\alpha$ and $\beta$ are the model  parameters then the likelihood can be expressed as \cite{PhysRevD.65.043506},
\begin{equation}\label{equ:like1}
\mathcal{L}(M_i)=\int
d\alpha \, p(\alpha|M_\text{i})\left(\int
d\beta \,  p(\beta|M_\text{i}) \, \mathcal{L}(\alpha,\beta) \right),
\end{equation}
where $\mathcal{L}(\alpha,\beta)$ is the likelihood of the combination of the model parameters $(\alpha, \beta)$ which can be computed  by marginalizing all model parameters.
Assuming the measurement errors are Gaussian, the likelihood function can be taken as \cite{Li:2009bn},
\begin{equation}\label{range}
\mathcal{L}(\alpha,\beta)\equiv e^{\frac{-\chi^2(\alpha,\beta)}{2}} ,
\end{equation}
where,
\begin{equation}\label{eqn:chi21}
\chi^2(\alpha,\beta)=\sum\left[\frac{{A_\text{k}}-A_\text{k}(\alpha,\beta)}{\sigma_\text{k}}\right]^2.
\end{equation}
Here ${A_\text{k}}$ is the observed value, $A_\text{k}(\alpha,\beta)$ is the corresponding theoretical value
and $\sigma_\text{k}$ is the uncertainty in the measurement of the observable.
We assume uniform prior information regarding the parameters such that they are lying in the range [$\alpha,\alpha+\Delta\alpha$]
and [$\beta,\beta+\Delta\beta$] respectively, then the prior probability of parameters can be taken as
$p(\alpha|M_\text{i})=\dfrac{1}{\Delta\alpha}$ and $p(\beta|M_\text{i})=\dfrac{1}{\Delta\beta}$ which are the simplest choice for these.
Hence the above equation~(\ref{equ:like1}) become
\begin{equation}\label{equ:like7}
\mathcal{L}(M_\text{i})=\dfrac{1}{\Delta\alpha}\dfrac{1}{\Delta\beta}\int_\alpha^{\alpha+\Delta\alpha}d\alpha^{\prime}\int_\beta^{\beta+\Delta\beta} d\beta^{\prime} \, exp[-\chi^2(\alpha^{\prime},\beta^{\prime})/2].
\end{equation}

Having the prior probabilities for the parameters, the marginal likelihood for a parameter, 
say $\alpha$ can be written as,
\begin{equation}\label{equ:like3}
\mathcal{L}_\text{i}(\alpha)=\dfrac{1}{\Delta\beta}\int_\beta^{\beta+\Delta\beta} d\beta^{\prime} \,  exp[-\chi^2(\alpha,\beta^{\prime})/2].
\end{equation}
Similarly marginal likelihood for $\beta$ can also be defined.
Knowing the likelihood of the models $M_\text{i}$ and $M_\text{j},$ the comparison between them can be performed 
by estimating Bayes factor
$B_\text{ij}$, which is the ratio of likelihood of the two
models,
\begin{equation}\label{equ:bayes}
B_\text{ij}=\frac{\mathcal{L}(M_{\text{i}})}{\mathcal{L}(M_{\text{j}})}.
\end{equation}
According to the conventional Jeffreys scale of reference in Bayesian analysis \cite{Jeffreys61} if the Bayes factor,
$B_\text{ij}<1$ then the model $M_\text{i}$ is not significant as compared to $M_\text{j}$. If $1<B_\text{ij}<3,$ there is 
evidence against $M_\text{j}$ when compared with $M_\text{i}$ but it is not worth more than a bare mention. For 
$3<B_\text{ij}<20,$ the evidence of $M_\text{i}$ against $M_\text{j}$ is not strong but definite. If $20<B_\text{ij}<150,$ the evidence is strong and on the other hand if $B_\text{ij}>150,$ evidence against $M_\text{j}$ is very strong \cite{Trotta:2005ar,Drell:1999dx,refId0}.

\section{The running holographic Ricci dark energy (rhrde) models}
In this section we are trying to give a relatively detailed description of the two types of the running holographic Ricci 
dark energy models that we have used. These are basically two component models, having running holographic Ricci dark energy and 
dark matter as the constituents.
The first model, model 1, having 
holographic Ricci dark energy and dark matter as cosmic components and the components are conserved together. In this model the dark
energy density is characterized with an additive constant, the bare cosmological constant which ensure the transition into the late acceleration epoch.
The second one, model 2, in which the interaction between the components 
is treated as 
non-gravitational in nature\cite{Guo:2007zk} and is accounted in a
phenomenological way.
We briefly describe both of these  models
before entering the Bayesian analysis \cite{Hee:2015eba,10.1093/mnras/stt1954,Saini:2003wq}.

\subsection{ Model 1}   
As mentioned previously this is the  model in which the Ricci dark energy is characterized 
with an additive bare cosmological constant
in its energy density. The density of dark energy has the form,
\cite{PhysRevD.87.023515,doi:10.1142/S0217732316500759}. 
\begin{equation}\label{eqn:den}
\rho_{\Lambda}(H,\dot{H})= 3\beta M_{\text{p}}^{2}(\dot{H}+2H^{2})+ M_{\text{p}}^{2}\Lambda_{0},
\end{equation}
where $M_{\text{P}}^2=\frac{1}{{8\pi \text{G}}}$ is the reduced Planck mass,  $\beta$ is the model parameter, $H$ is the
Hubble parameter and $\dot{H}$ is its derivative with respect to cosmic time. This is running in the sense that its
equations of state is fixed, $\omega =-1$ while the density is varying as the universe evolves.

The original idea of running dark energy, also called running vacuum energy, were proposed \cite{doi:10.1142/S0217751X16300350,Sol__2013} based on the quantum field theory on curved space time in order to alleviate the cosmological constant problem\cite{Shapiro:2009dh}. 
Following these idea and using the Renormalization group (RG) approach, the running vacuum 
energy has been proposed as\cite{Lima:2012mu,Sol__2013},
\begin{equation}\label{rv1}
\rho_{\Lambda}= \frac{3}{8\pi G} \left(c_0 + \nu H^2 + ......\right),
\end{equation}
where $c_0$ often takes the role of a bare cosmological constant and 
$\nu=\frac{1}{6\pi}\sum_\text{i} b_\text{i}
\frac{M_\text{i}^2}{M_\text{P}^2},$ a dimensionless parameter, where $b_\text{i}$ is the coefficients computed form the loop contributions of the 
fields with masses $M_\text{i}.$ The numerical value of the parameter $\nu$ is very small\cite{Sol__2013}, however it gives the running status to the vacuum 
energy. A more general form of Equation~(\ref{rv1}) for the current universe  
can be expressed as a power series of the 
Hubble function(${H}^2$) and its derivatives($\dot{{H}}$)\cite{Sol__2017,Sola:2016vis} as, 
\begin{equation}\label{rv2}
\rho_{\Lambda}=3M_{\text{p}}^2(c_0+\nu H^2+\frac{2}{3}\alpha \dot{{H}})+\mathcal{O}(H^4),
\end{equation}
which has one more free parameter $\alpha,$ the value of which is comparable to that of $\nu.$ The $\mathcal{O}(H^4)$ terms in the equation(\ref{rv2}) are irrelevant for the study of the current universe, but are essential for the correct account of the inflationary epoch. In 
reference\cite{Gomez-Valent:2014rxa}, authors have contrasted the model with the recent observational data on type Ia supernovae(SN1a), 
the Cosmic Microwave Background(CMB) and the Baryonic Acoustic Oscillations (BAO) and it effectively 
supports a slowly varying `cosmological constant'. 
Using the data on expansion, structure formation, BBN(Big Bang Nucleosynthesis) and CMB observables authors in the 
reference\cite{Sol__2015} shows that $\Lambda$CDM model, which support a rigid cosmological constant, is currently disfavored 
at 3$\sigma$ level and on the other hand these data actually favoring 
a slowly running vacuum. 
This model could able to fit the combined cosmological data (SNIa+BAO+H(z)+LSS+BBN+CMB) significantly better than the 
concordance $\Lambda$CDM model\cite{Sol__2017}. 

The running dark energy we have used is generically of similar form, but basically follows from the holographic 
principle, according 
to which the total vacuum energy in a region of size $L$ should not exceed the mass of a black hole  of the same size, i.e. 
$L^3  \rho_\text{vac} \leq L M_\text{P}^2$ where $\rho_\text{vac}$ is the quantum zero point energy density.
Based on this, Li \cite{Li:2004rb} proposed 
the holographic dark energy in cosmology as $\rho_\text{vac}=3 c^2 M_{\text{P}}^2 L^{-2}$ where $c^2$ is a dimensionless constant and $L$ 
is called the
IR cut off corresponding to a relevant cosmological length scale. Following this Gao\cite{Gao:2007ep} proposed what is called as 
the holographic Ricci dark energy, $\rho_{\Lambda}=-\frac{\alpha}{16\pi}R$ where 
$R=-6\left(\dot H+ 2H^2\right)$ is the Ricci scalar for a flat universe. This model of dark energy has the drawback that, it will not implies 
a transition into a late accelerating universe. In order to alleviate this issue, a bare cosmological constant has been added to the initial 
form of the holographic Ricci dark energy, it then takes a more general form as given in equation (\ref{eqn:den}). The model parameter 
$\beta$ in this case can be expressed in terms of the original RG running parameter $\nu$ as $\beta \sim \frac{\nu}{2}
\sim \frac{2\alpha}{3}$ \cite{doi:10.1142/S0217732316500759}.

Having equation(\ref{eqn:den}) for the running vacuum energy,
the  conservation law followed by the major cosmic components is (taking account of radiation also),
\begin{equation}\label{eqn:con1} 
\dot{\rho}_{\text{m}}+\dot{\rho}_{\text{r}}+\dot{\rho}_{\Lambda}+3H(\rho_{\text{m}}+\frac{4}{3}\rho_{\text{r}})=0,
\end{equation}
where $\rho_m,\rho_r$ are the matter and radiation densities respectively and the over dot represents their derivatives
with respect to cosmic time.
From the Friedmann equation the Hubble  parameter, $h=H/H_0$  takes the form \cite{doi:10.1142/S0217732316500759},
\begin{equation}\label{eqn:h}
h^{2}=\frac{\Omega_{m_{0}}}{\xi_{\text{m}}}e^{-3\xi_{\text{m}}x}+\Omega_{r_{0}}e^{-4x}+\frac{\Lambda_0}{3(1-2\beta)H_{0}^{2}},
\end{equation}
where
$\Omega_{m_{0}}=\frac{\rho_{m_{0}}}{3 M_{\text{p}}^{2}H_{0}^{2}}$,
$\Omega_{r_{0}}=\frac{\rho_{r_{0}}}{3 M_{\text{p}}^{2}H_{0}^{2}}, \, H_0$ is the present value of the Hubble parameter and
$\xi_{m}=\frac{(1-2\beta)}{(1-\frac{3}{2}\beta)}.$  The variable $x=ln \, a.$ The first term in Equation~(\ref{eqn:h})
corresponds to matter density,
second term represents the energy density of radiation and the third term corresponds to the bare cosmological constant. This equation implies an upper limit on the model parameter, $\beta < 1/2.$ In the limit $a\to \infty$ the Hubble
parameter becomes,   $h^2 \to
\frac{\Lambda_0}{3(1-2\beta)H_{0}^{2}},$ a constant which corresponds to an accelerating epoch universe dominated 
by the bare cosmological constant and it is this bare cosmological constant is the sole reason for the 
acceleration\cite{doi:10.1142/S0217732316500759,PhysRevD.87.023515}.\\

The background evolution of this model has been discussed in a previous work\cite{doi:10.1142/S0217732316500759}, where it was shown that, due to the 
presence of the additive constant in the density, the model predicts a transition into a late accelerating epoch at around 
of $z_T = 0.7.$ It could also be interesting to note that 
the matter density get modified as, $\Omega_m=\Omega_{m_0}a^{-3\xi_m}$ whereas the radiation satisfies 
the conventional behavior as $\Omega_r=\Omega_{r_0}a^{-4}$\cite{doi:10.1142/S0217732316500759}.  This implies that the radiation component can be 
omitted safely in this model to describe the evolutionary history of the universe.

\subsection{ Model 2} 
This model, what we called also as interacting rhrde, has got two basic differences from model 1. The first one is that
the dark energy density is of
the same form as in the previous case, but devoid of the additive constant. In addition, the interaction between
the dark sectors is taken care of by a phenomenological term,
$Q=3bH\rho_\text{m},$ where $b$ is the coupling constant and $\rho_m$ is density of dark matter
\cite{PhysRevD.81.103514,PhysRevD.83.063515,PhysRevD.78.123503,ZIMDAHL2001133}. The conservation equations then follows as,
\begin{equation}
\begin{split}
\dot{\rho}_{\text{hrde}}+3H(\rho_{\text{hrde}}+P_{\text{hrde}})=-Q,\\
\dot{\rho}_{\text{m}}+3H(\rho_{\text{m}}+P_{\text{m}})=Q.
\end{split}
\end{equation}
In a previous work we had analyzed the background evolution of the universe in this model\cite{George:2018myt} and have shown 
that even without the additive constant in the dark energy density, the model predicts a transition into the late accelerating epoch 
and is basically due to the presence of phenomenological interaction term. On
solving the Friedmann equation,
\begin{eqnarray}
3H^2= \rho_m + \rho_{\text{hrde}},
\end{eqnarray}
for the Hubble parameter we get,
\begin{equation}\label{equ:hub}
\begin{split}
h^{2}=\frac{\Omega_{m_0}}{1-b} e^{-3(1-b)x} -\frac{1}{3}\left(\frac{2\Omega_{hrde_0}}{\beta}+3\Omega_{m_0}-4\right)e^{-3x}+\\
\left(\frac{2\Omega_{hrde_0}}{3\beta}-\frac{b}{1-b}\Omega_{m_0}-\frac{1}{3}\right),
\end{split}
\end{equation}
where
$ \Omega_{hrde_0}=\frac{\rho_{hrde_0}}{3M_{\text{p}}^2H_0^2},$
with $\rho_{hrde_0}$ as the present value of the dark energy density.
The variable, $x=ln a,$ where $a$ is the scale factor of expansion. The Hubble parameter has the expected asymptotic
properties. As $x\to -\infty$(equivalently $a\to 0$) the first two terms in equation~(\ref{equ:hub}) dominate
which implies a prior deceleration phase. As $x \to +\infty$ (equivalently $ a \to \infty$) Hubble parameter tends to a
constant which corresponds to the end de Sitter phase. 

The transition into the late accelerating epoch is found to occur at a redshift $z_T = 0.71$ for SN1a(307)+\\CMB(WMAP7)+BAO
data and is sightly higher, $z_T = 0.74$, for SN1a(307)+CMB (Planck2013)+BAO data and is in good
agreement with the observational result\cite{Alam:2004jy}. 
We have also analyzed model with statefinder diagnostic, and found that trajectory of the model in the state finder 
parameter plane, i.e. $(r, s)$ plane (where $r$ is the jerk and $s$ is the snap parameter), is restricted in the region 
corresponding to $r < 1$ and $s > 0$, which implies the
quintessence nature of the running vacuum.
Moreover the dynamical system analysis of the model with a suitably constructed
phase-space shows that the late acceleration phase is asymptotically stable\cite{George:2018myt}

\subsection{Data analysis for model parameters}
In this section we obtain the 
parameters of both model 1 and model 2. Model 1 have the following as the 
parameters, $\beta, \Omega_{m0}, \Omega_{\Lambda}$ and $H_0$ (see equations (\ref{eqn:den}) and (\ref{eqn:h})). In the case of model 2
there exists one more free parameter and that is $b,$ (see equation(\ref{equ:hub}))
characterizing the interaction between the dark sectors. Our aim here to 
extract the magnitude of all parameters 
using the 
observational data. These model parameters, which will give us a reasonable idea about the evolutionary status of the universe in these models, are evaluated using the $\chi^2$ minimization technique\cite{Ranjit:2014fra,Paul:2014kza,Thakur:2017syt}.  
For the computation we have used supernova 580 data from the SCP "Union2.1" SN Ia compilation \cite{2012ApJ...746...85S}, 
$H(z)$ data \cite{Farooq_2017}, CMBR data from Planck2015 \cite{Ade:2015rim,PhysRevD.89.063510} and BAO (Baryon Acoustic 
Oscillation)
data from Sloan Digital Sky Survey(SDSS) \cite{Eisenstein:2005su,PhysRevD.74.123507}. We repeat the computation
by replacing 580 supernovae data set with
the 307 data from Union compilation \cite{Kowalski:2008ez}. The 580 data set consists of more data points from the low
redshift observations. It has been noted that  interference with the background in obtaining the low redshift data is
relatively high \cite{2004PhT....57f..19S} and also Kolmogorv-Smirnov analysis \cite{Bianco_2011} has shown that 307
data set has relatively larger, i.e. 85$\%$ probability of having originated from a common distribution.
The theoretical distance modulus for a redshift $z_\text{i}$ is,
\begin{equation}\label{equ:mu}
\mu_\text{t}(\beta,b,H_0,z_\text{i})=5\log_{10}\left[\frac{d_{L}(\beta,b,H_0,z_\text{i})}{Mpc}\right]+25,
\end{equation}
where
$d_{L}$ is its luminosity distance.
The $\mu_\text{t}$ calculated for a given redshift is to be compared with the observational data for the same redshift for 
obtaining $\chi^2$ using equation~(\ref{eqn:chi21}) by replacing  $A_\text{k}$ with the distance modulus $\mu.$

We have used Hubble parameter data
\cite{Farooq_2017}, containing 38 samples in the red shift range $0.07\leq z\leq 2.36.$ The $\chi^2$ has 
been obtained using equation~(\ref{eqn:chi21}), in which we replace $A_\text{k}$ with $H.$
In using Cosmic Microwave Background (CMB) data \cite{Bond:1997wr},
the shift parameter $\mathcal{R},$ is taken as the observable instead of $A_\text{k}$ in equation~(\ref{eqn:chi21}), defined as,
\begin{equation}\label{equ:param2}
\mathcal{R}=\sqrt{\Omega_{{m}}}\int_{0}^{z_{2}}\frac{dz}{h(z)}.
\end{equation}
Here $z_{2}$ is the red shift at the last scattering surface. From Planck 2015 data, $z_2=1089.9$ and $\mathcal{R}=1.7382 \pm 0.0088$  \cite{Ade:2015rim,PhysRevD.89.063510}.
For Baryon Acoustic Oscillation(BAO) data, the observable used in equation~(\ref{eqn:chi21}) is the acoustic parameter $ \mathcal{A},$
\begin{equation}\label{equ:param1}
\mathcal{A}=\frac{\sqrt{\Omega_{\text{m}}}}{h(z_{1})^{\frac{1}{3}}}\left(\frac{1}{z_{1}}\int_{0}^{z_{1}}\frac{dz}{h(z)}\right)^{\frac{2}{3}}.
\end{equation}
Here
$z_{1} = 0.35$ is
the redshift where the signature of the peak acoustic
oscillations has been measured \cite{PhysRevD.74.123507}.
According to the SDSS data  the observational value of the acoustic parameter for flat
universe corresponding to the same redshift is, $\mathcal{A} = 0.484 \pm 0.016$ \cite{10.1111/j.1365-2966.2011.19592.x}.

We did the $\chi^2$ (defined in equation(\ref{eqn:chi21})) minimization process 
to extract the parameters $\beta$, $b$, $H_0, \Omega_{m0}$ and $\Omega_{\Lambda}.$ 
We obtained the parameters for the data combinations SNIa(307)+BAO+CMB, SNIa(307)+BAO+CMB+Hubble data, SNIa(580)+BAO+CMB and
SNIa(580)+BAO+CMB+Hubble data which leads to a total $\chi^2$ of the form,
$\chi^2(\beta,b,H_0,\Omega_{m0},\Omega_{\Lambda})=\chi^2(\beta,b,H_0,\Omega_{m0},\Omega_{\Lambda})_{\text{SNIa}}+\chi^2(\beta,b,H_0,\Omega_{m0},\Omega_{\Lambda})_{\text{CMB}}+\chi^2(\beta,b,H_0,\Omega_{m0},\Omega_{\Lambda})_{\text{BAO}}+
\chi^2(\beta,b,H_0,\Omega_{m0},\Omega_{\Lambda})_{\text{Hubble data}}$.
The best fit values of the parameters for model 1  and model 2 are estimated with 1$\sigma$ level of correction and the results are shown in table~\ref{table:1a} and table~\ref{table:2a} respectively. 
\begin{table}
	\centering
	\caption{The table gives value of the parameter $\beta$ in model 1 for different data combinations. In the table data1 refers to the combination SNIa(307)+CMB+BAO, data2 refers to SNIa(307)+CMB+BAO+Hubble data, data3 refers to SNIa(580)+CMB+BAO, data4 refers to SNIa(580)+CMB+BAO+Hubble data}
	\label{table:1a}
	\begin{scriptsize}
		\begin{tabular}{lcccccr} 
			\hline
			Parameter & data1 &  data2 & data3 & data4 \\[0.05in]
			\hline
			$\beta$  & $0.000055_{-0.006}^{+0.006}$ & $0.0046_{-0.006}^{+0.005}$& $0.0012_{-0.004}^{+0.005}$& $0.0039_{-0.005}^{+0.005}$   \\[0.05in]
			$\Omega_{m_0}$ & $0.2760_{-0.006}^{+0.007}$ & $0.2617_{-0.007}^{+0.008}$ & $0.2706_{-0.007}^{+0.007}$& $0.2649_{-0.006}^{+0.007}$\\[0.05in]
			$\Omega_{\Lambda}$ &$0.7095_{-0.015}^{+0.015}$ &$0.7153_{-0.016}^{+0.016}$  &$0.7075_{-0.014}^{+0.014}$ &$0.7171_{-0.014}^{+0.015}$ \\[0.05in]
			$H_0$ & $70.63_{-0.52}^{+0.55}$ &$70.78_{-0.50}^{+0.51}$  &$70.78_{-0.46}^{+0.47}$ &$70.45_{-0.44}^{+0.46}$ \\[0.05in]
			$\chi^2_{min}$& 313.62&355.41 &562.24 &607.71 &\\[0.05in]
			$\chi^2_{d.o.f}$ &$1.02$ &$1.04$ &$0.97$ &$0.99$\\
			\hline
		\end{tabular}
	\end{scriptsize}
\end{table}
\begin{table}
	\centering
	\caption{The value of the parameter $\beta$ and b for different data sets of model 2 are given. In the table data1 refers to the combination SNIa(307)+CMB+BAO, data2 refers to SNIa(307)+CMB+BAO+Hubble data, data3 refers to SNIa(580)+CMB+BAO, data4 refers to SNIa(580)+CMB+BAO+Hubble data}
	\label{table:2a}
	\begin{scriptsize}
		\begin{tabular}{lcccccr} 
			\hline
			Parameter & data1 &  data2 & data3 & data4 \\[0.05in]
			\hline
			$\beta$  &$0.4583_{-0.003}^{+0.002}$ & $0.4625_{-0.004}^{+0.003}$&$0.4597_{-0.004}^{+0.004}$&$0.4639_{-0.003}^{+0.003}$    \\[0.05in]
			b&$0.0055_{-0.006}^{+0.006}$&$0.0063_{-0.007}^{+0.008}$&$0.0054_{-0.007}^{+0.007}$&$0.0059_{-0.006}^{+0.006}$\\[0.05in]
			$\Omega_{m_0}$ & $0.2947_{-0.007}^{+0.007}$ & $0.2768_{-0.008}^{+0.008}$ & $0.2918_{-0.007}^{+0.007}$& $0.2759_{-0.006}^{+0.007}$\\[0.05in]
			$\Omega_{\Lambda}$ &$0.7078_{-0.013}^{+0.014}$ &$0.7279_{-0.014}^{+0.014}$ &$0.7155_{-0.013}^{+0.013}$ &$0.7044_{-0.014}^{+0.015}$ \\[0.05in]
			$H_0$ & $70.03_{-0.53}^{+0.54}$ &$70.28_{-0.53}^{+0.53}$  &$70.06_{-0.47}^{+0.49}$ &$70.03_{-0.51}^{+0.50}$ \\[0.05in]
			$\chi^2_{min}$& 312.41&356.42 &562.22 &608.76 &\\[0.05in]
			$\chi^2_{d.o.f}$ &$1.03$ &$1.04$ &$0.97$ &$0.99$\\
			\hline
		\end{tabular}
	\end{scriptsize}
\end{table}

For model 1, the value of the parameter $\beta \sim 0.000055$ for SNIa(307)+CMB+BAO data and found to be almost the same around $\beta \sim 0.001$ for all the other data combinations.
In the case of model 2, the parameter $\beta$ is found to be 
almost hundred times large, around $\beta \sim 0.46,$ however the values 
remains almost the same for all data combinations. 
The interaction parameter $b$ of model 2 is found to be around $b \sim 0.006$ for all data combination. 
The increase in the parameter $\beta$ for model 2 implies that with the phenomenological interaction the strength of
dark energy has increased substantially.

\begin{figure}
	\centering
	\includegraphics[scale=0.55]{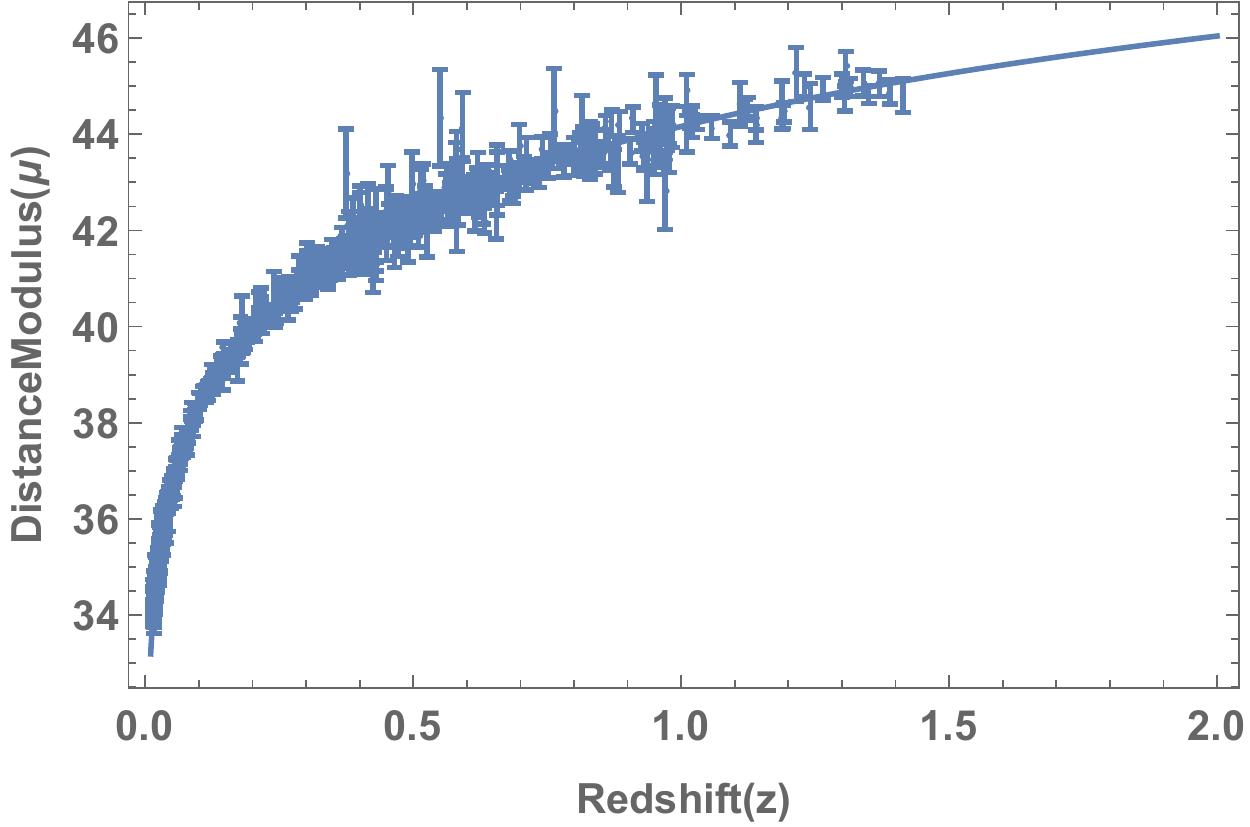}
	\caption{Plot of distance modulus ($\mu$) versus red shift (z) of model 1.}\label{figa}
\end{figure}

\begin{figure}
	\centering
	\includegraphics[scale=0.55]{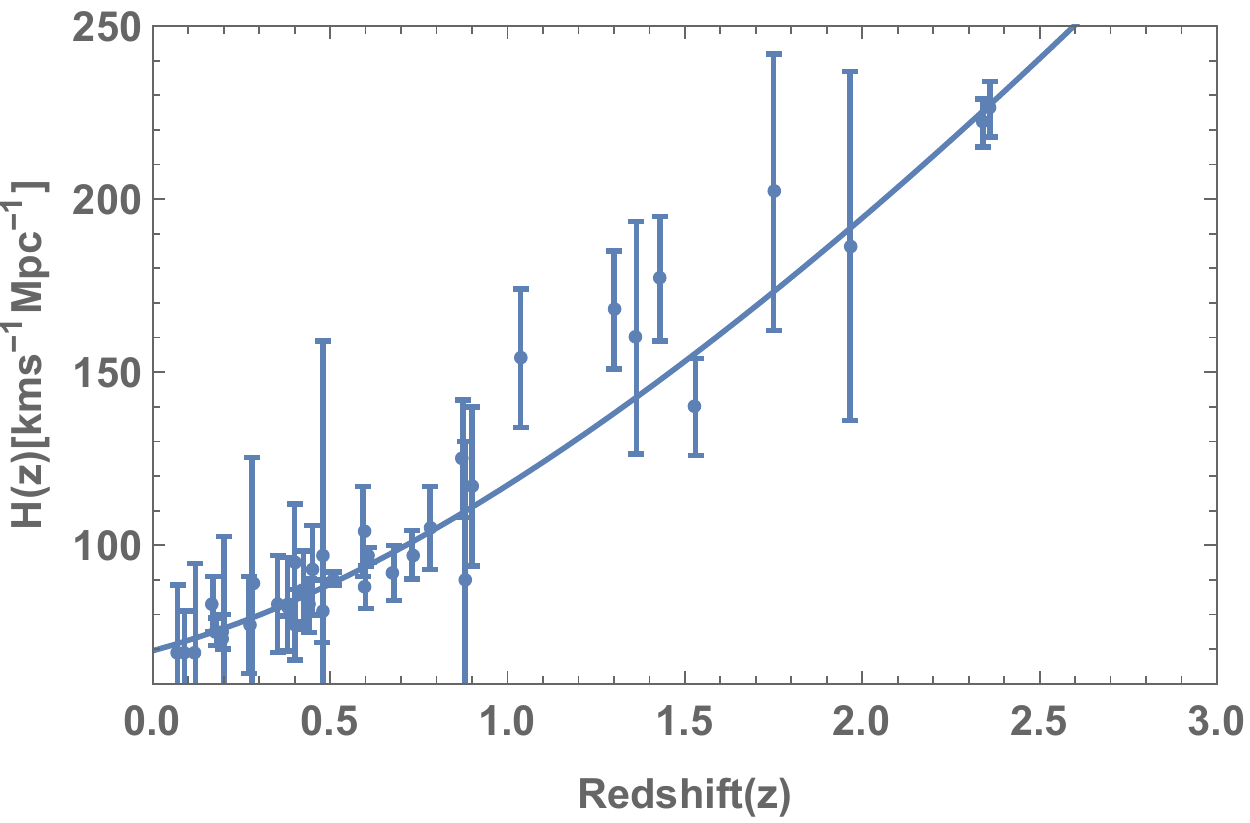}
	\caption{Plot of Hubble function (H) versus red shift (z) of model 1.}\label{figc}
\end{figure}

\begin{figure}
	\centering
	\includegraphics[scale=0.55]{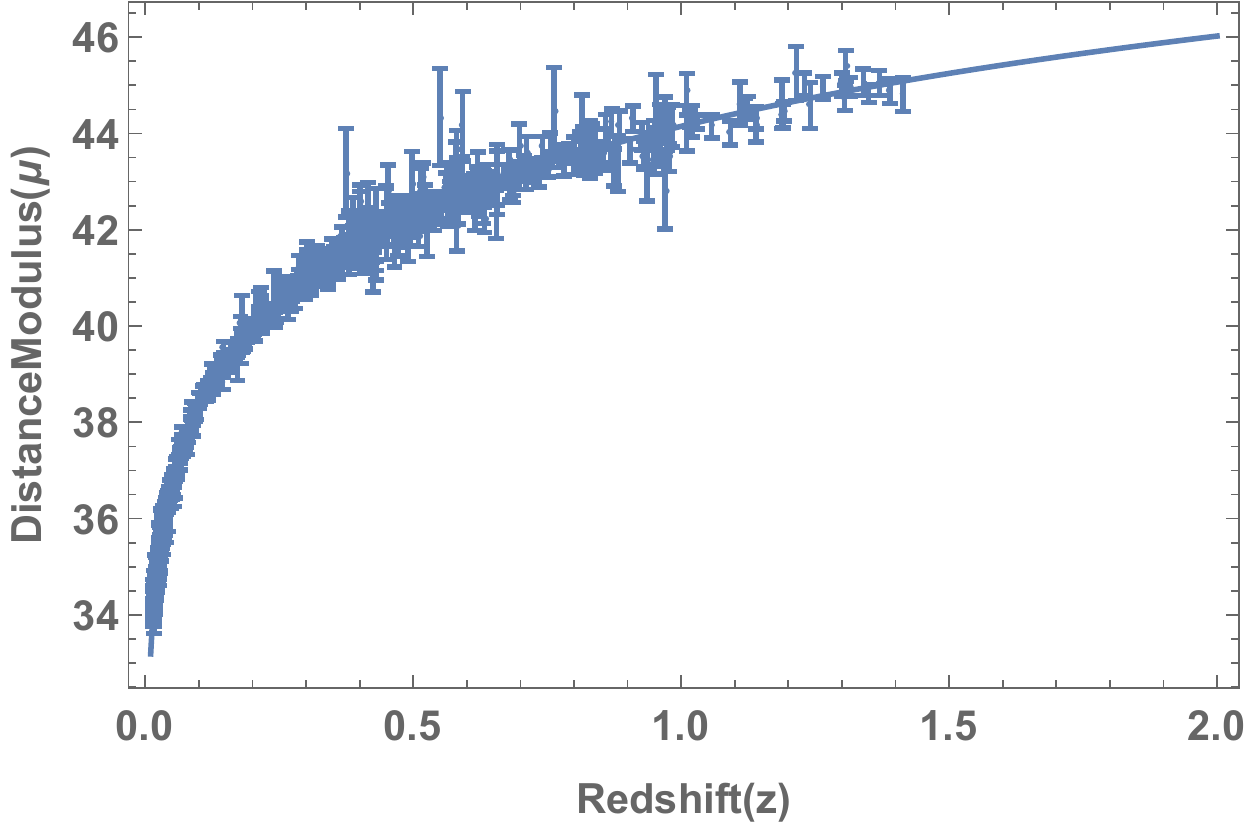}
	\caption{Plot of distance modulus ($\mu$) versus red shift (z) of model 2.}\label{figb}
\end{figure}

\begin{figure}
	\centering
	\includegraphics[scale=0.55]{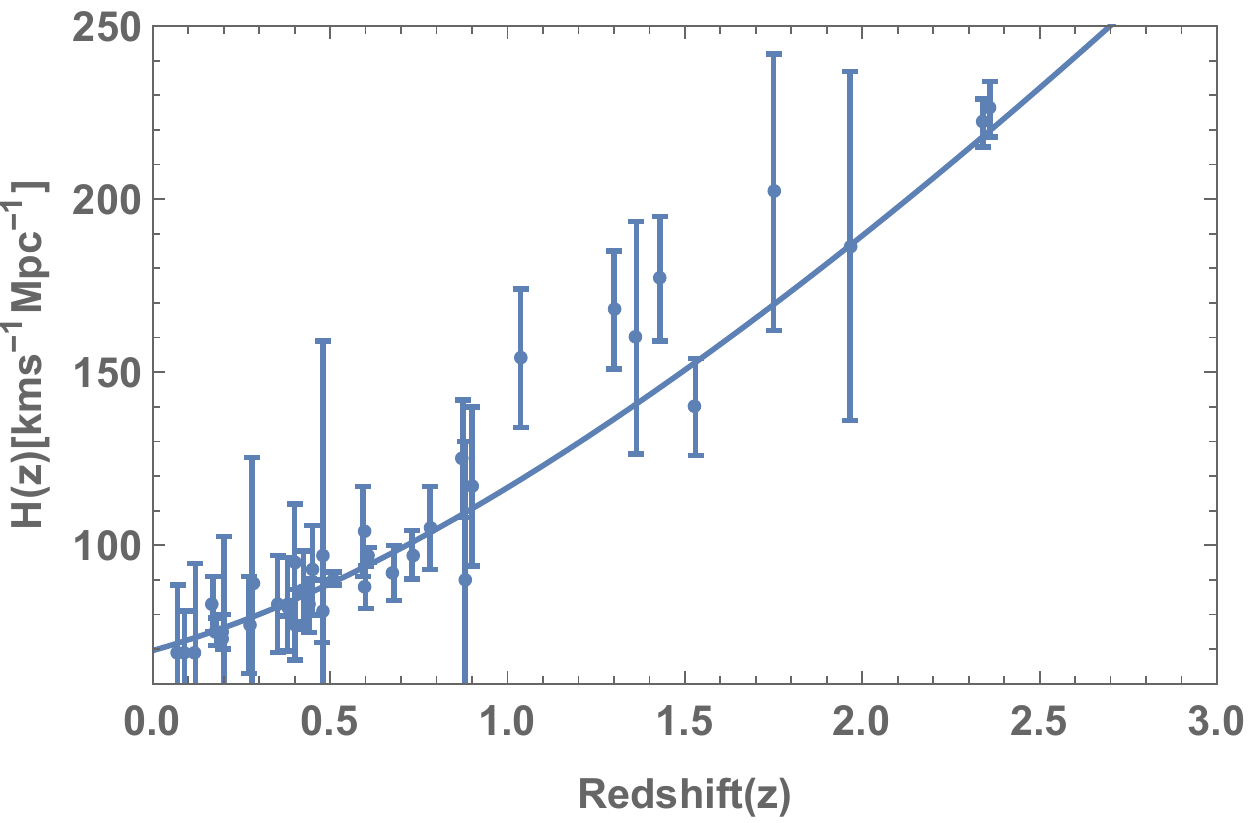}
	\caption{Plot of Hubble function (H) versus red shift (z) of model 2.}\label{figd}
\end{figure}

Now we will briefly discuss 
the behavior of the models in predicting the distance modulus and Hubble parameter  
with the best-fit values of the model parameters  corresponding to the latest data, the data combination,
SNIa(580)+BAO+CMB+Hubble data. The distance modulus for 
different red shifts using equation(\ref{equ:mu}) and the evolution of the 
Hubble parameter using the relations (\ref{eqn:h}) for model 1 and (\ref{equ:hub}) for model 2 are computed by adopting 
the above mentioned parameter values. 
The theoretical distance modulus ($\mu$) for respective red shifts  
is compared with the corresponding observational data in figures(\ref{figa}) and (\ref{figb})
whereas the evolution of the Hubble parameter with red 
shift in comparison with the observed values are shown in figures(\ref{figc}) and (\ref{figd}).  
The solid line in the figures, represent the theoretical curve for the best fit values determined 
by the joint analysis using SNIa(580)+CMB+BAO+stern data, shows good agreement with the observational data. 


\section{Bayesian analysis for model 1}

We will now calculate the likelihood of the model 1 \cite{doi:10.1142/S0217732316500759}. This model have rhrde 
and dark mater as the cosmic components and are satisfying a common conservation law.  The 
basic inputs in computing the Bayes factor are the priors of the parameters, $\beta, \Omega_{m0}, \Omega_{\Lambda} \,  
\textrm{and} \, H_0.$ We choose flat priors in advance without considering the data.
The Bayes factor calculation can be done by marginalizing over all these parameters. 
In the following we will describe the prior ranges chosen 
for different parameters.
\begin{enumerate}
	\item Parameter $\beta:$
	We adopt a prior for $\beta$ in the range 0 - 0.5.
	This can be justified as follows.
	In model 1 we have considered the running holographic Ricci dark energy (rhrde) density as given by equation(\ref{eqn:den}).
	This equation clearly indicate that the parameter 
	$\beta$ must have a value greater than zero for 
	keeping its identity as running vacuum. 
	On solving the Friedmann equation 
	(along with continuity equation) for Hubble parameter, $h=H/H_0$ we get\cite{doi:10.1142/S0217732316500759} the solution as given in equation(\ref{eqn:h}). 
	On substituting this solution 
	in equation (\ref{eqn:den}) lead to  a general expression for the running vacuum as,
	\begin{equation}\label{eqn:tot de dens}
	\rho_{\Lambda}=\frac{3\beta M_{p}^{2}H_{0}^{2}}{2-4\beta}\Omega_{m_{0}}e^{-3\xi_{m}x}+\frac{M_{p}^{2}\Lambda_0}{1-2\beta}.
	\end{equation}
	This indicate that 
	the parameter has upper limit corresponding to $\beta < 1/2$ for $\rho_{\Lambda}$ being the running vacuum with positive definite 
	magnitude. This justifies the range of $\beta$ as $0 - 0.5.$ 
	\item Parameter $\Omega_{m0}:$
	The prior for $\Omega_{m0}$ chosen in the present work is 0.1$\leq\Omega_{m0}\leq$0.7\cite{Ryan:2019uor}. This range has been chosen in
	comparison with the almost similar ranges that used in many references, for instance: 0$\leq\Omega_{m0}\leq1$\cite{Weller:2001gf,Gupta:2010jp}, 
	0.2$\leq\Omega_{m0}\leq$1.0\cite{Lima:2011ye}, 0.01$\leq\Omega_{m0}\leq$1\cite{Ryan:2018aif},0.01$\leq\Omega_{m0}\leq$0.6\cite{Tripathi:2016slv}. 
	The authors in \cite{Weller:2001gf} used the above prior $0\leq\Omega_{m0}\leq1$ to show how an independent measurement 
	of $\Omega_{m0}$ can allow SNAP(SuperNovae Acceleration Probe) to probe the evolution of the equation of state and thus 
	allowing to discrimination among a larger class of proposed dark energy models. In the reference\cite{Tripathi:2016slv}, 
	using the prior 0.01$\leq\Omega_{m0}\leq$0.6, the authors have deduced strong limits on variation of dark energy density 
	with redshift by using the observational data,
	SNIa, BAO and Hubble parameter measurements. 
	\item Parameter $\Omega_{\Lambda}:$ 
	This is the present value of the mass density parameter of rhrde.
	We have adopted a prior range, $0.2\leq\Omega_{\Lambda}\leq1.$  We have chosen this range in comparison with  
	references; 0$\leq\Omega_{\Lambda}\leq$1\cite{Ryan:2018aif,Marinoni:2007vn}, where it is used to constrain the present 
	value of the mass parameter of dark energy and 0$\leq\Omega_{\Lambda}\leq$2\cite{Ettori:2002pe},  
	used for constraining the both dark energy density and dark matter density.
	\item Parameter $H_0:$
	For the present value of the Hubble parameter, we adopt a range, 65 - 73 km$s^{-1}Mpc^{-1}.$
	Now let us explain this range adopted for the computation of likelihood.  
	The recent Planck observations\\\cite{2018arXiv180706209P} measured a 
	value of $H_0$ as $67.4\pm 0.5 $\, km $s^{-1}Mpc^{-1}.$ On the other hand a higher value, $73.24\pm1.74$ km$s^{-1}Mpc^{-1},$ 
	was obtained for $H_0$ from the local 
	expansion rate estimation\cite{Riess:2016jrr}. The authors in the reference\cite{Tripathi:2016slv} consider a prior range for $H_0$ as 65-75 km$s^{-1}Mpc^{-1}$ which is close to selected prior range. Considering all these, we choose a concordance prior range for $H_0,$ as 
	mentioned above.
\end{enumerate}

The basic
equation for computing the likelihood is
equation~(\ref{equ:like7}), which is expressed for two general parameters
$\alpha$ and $\beta.$ For model 1, we extended this equation analogously for four parameters, $\beta, \Omega_{m0}, \Omega_{\Lambda} \, \textrm{and} \, H_0.$ 
The term $\Delta$ of each parameter is the width of the prior range of the respective parameters. 
The computations has been performed for four different combinations of the data , 
1.SNIa(307)+CMB+BAO, 2.SNIa(307)+CMB+BAO\\+Hubble data, 3.SNIa(580)+CMB+BAO and 
also 4.SNIa(580)+CMB+BAO+Hubble and the values are given in table~\ref{table:1}. The difference between the data sets 1 $\&$ 2 and 3 $\&$ 4 is that in the former set we have used 
307 data compositions of type IA supernovae while the later set the supernovae data used is latest 580 set.

\begin{figure}
	\centering
	\includegraphics[scale=0.55]{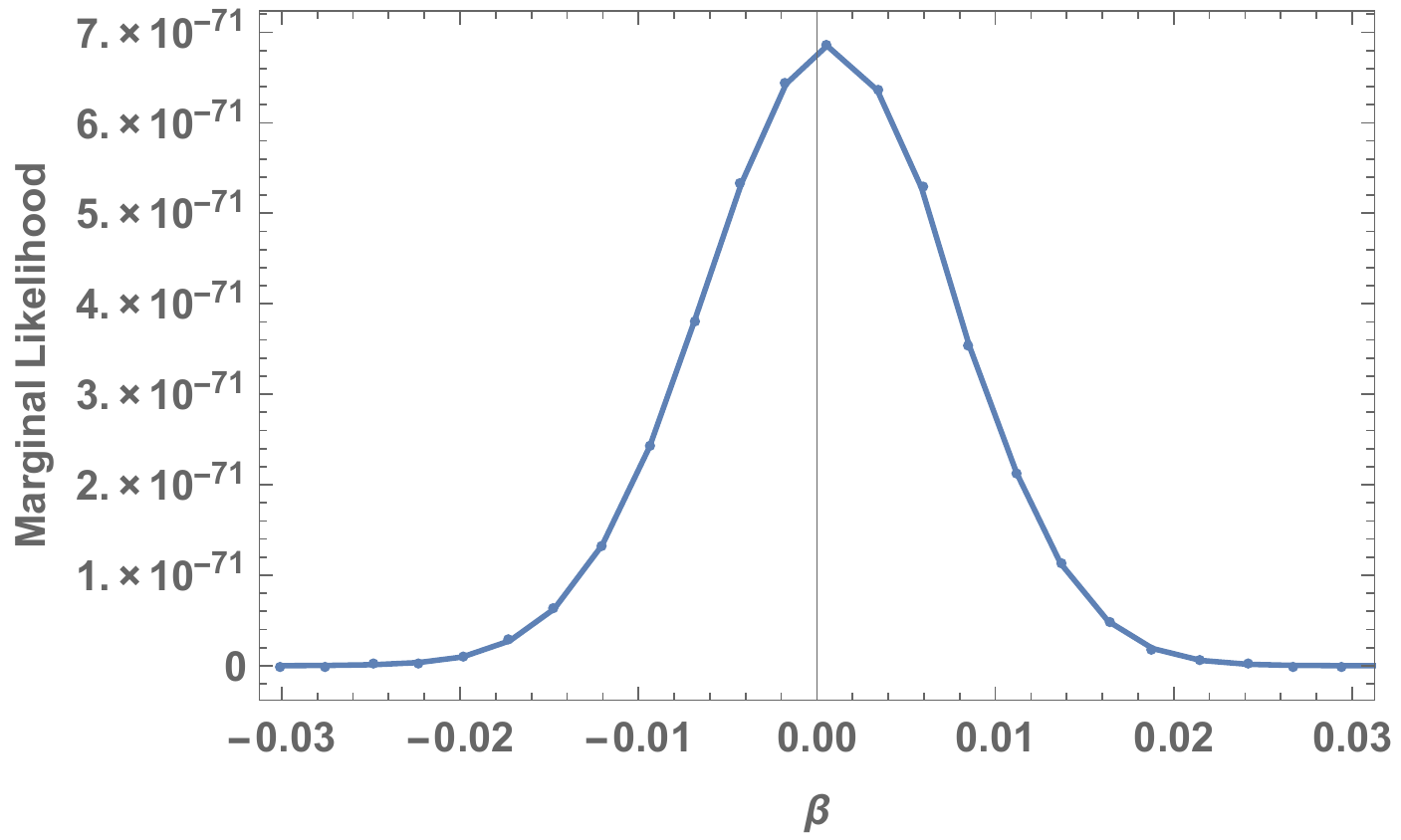}
	\caption{The plot shows the marginal likelihood of the model parameter $\beta$ in model 1 for SNIa(580)+CMB+BAO 
		data set.}\label{fig:f01}
\end{figure}

The robustness of the likelihood calculation can be exposed by calculating the marginal likelihood. 
For instance the marginal likelihood for the model parameter $\beta$ can be obtained by integrating over all other 
parameters except $\beta$ as in relation~(\ref{equ:like3}). 
The resulting function is then plotted against $\beta$ and is found to have a
Gaussian shape, see  Fig.~(\ref{fig:f01}), which is peaked around $\beta \sim 0.000055.$ This is the same value we have obtained for 
$\beta$ with the same data by using the method of $\chi^2$ minimization in the previous section.

For obtaining the Bayes factor of the model, we have to evaluate the likelihood of the standard
$\Lambda$CDM also for the same data sets. The standard model is a two component model with dark matter 
($\Omega_{\text{m}}=\frac{\rho_{\text{m}}}{3 H^2}$) and cosmological constant 
$\Lambda$ ( $\Omega_{\Lambda}=\frac{\rho_{\Lambda}}{3 H^2}$) as the constituents.
There are three parameters in $\Lambda$CDM model, $\Omega_{m_0}, \Omega_{\Lambda}$ and $H_0.$ 
We have computed the likelihood, $\mathcal{L}(M_\text{j})$ of the $\Lambda$CDM with same priors for the parameters using the 
corresponding relation~(\ref{equ:like7}).
The Bayes factor for the model 1 in comparison to the standard $\Lambda$CDM model 
is then 
evaluated using the
expression~(\ref{equ:bayes})
and the results are summarized in table~(\ref{table:1}).
\begin{table}
	\centering
	\caption{The likelihood and Bayes factor for different data sets of $\Lambda$CDM and model 1 are given in the table. Here data1 refers to the combination SNIa(307)+CMB+BAO, data2 refers to SNIa(307)+CMB+BAO+Hubble data, data3  refers to SNIa(580)+CMB+BAO, data4 refers to SNIa(580)+CMB+BAO+Hubble data}\label{table:1}
	\begin{tabular}{lcccccr}
		\hline
		Data & Likelihood of  & Likelihood of  & Bayes Factor\\ [0.2ex] 
		&   $\Lambda$CDM($\mathcal{L}(M_j)$) &   model 1($\mathcal{L}(M_i)$)		&   $B_{ij}=\frac{\mathcal{L}(M_i)}{\mathcal{L}(M_j)}$&          \\
		\hline
		data1 & 6.8607$\times 10^{-71}$& 1.10706$\times 10^{-72}$ &0.0161  \\ 
		data2 & 2.82795$\times 10^{-80}$ & 5.17108$\times 10^{-80}$& 0.0182 \\
		data3 & 2.51447$\times 10^{-125}$ &4.33674$\times 10^{-127}$ & 0.0172 \\
		data4 & 2.60201$\times 10^{-135}$ & 4.68398$\times 10^{-137}$&0.0180 \\
		\hline
	\end{tabular}
\end{table}

According to table~(\ref{table:1}), the Bayes factor of model 1 with respect to the standard $\Lambda$CDM for all data combinations 
are below 1. For data 1 the Bayes factor is 0.0161, while for data 2, which consists of Hubble parameter data in addition, 
the Bayes factor is increased slightly to 0.0182. 
For the full data set, SNIa(580)+CMB+BAO+Hubble the Bayes factor is 0.0180.  According to Jefferys 
scale if the value of Bayes factor is in the range $B_\text{ij}<1,$ then 
the evidence of model 1 is very weak. This in turn implies that 
the evidence of the standard 
$\Lambda$CDM model against the model 1 is very strong. To be specific, compared to
the rhrde model with a bare cosmological constant in the dark energy density, the preference 
for the standard $\Lambda$CDM model is strong. 

\section{Bayesian analysis for model 2 }

Now we will go for the Bayesian analysis of the model 2, 
in which the interaction between the dark sectors is accounted with a
phenomenological term \cite{George:2018myt}. In contrast to the previous model
the late acceleration is caused by this interaction term instead of the constant additive term in the dark energy density.
Here also we have used the four data set combinations for the analysis as in the previous case. 
This model contains the same parameters like model 1 and in addition to that it have a parameter $b$ characterizing the interaction  
between the dark sectors. 

We use the same priors for $\beta, \Omega_{m_0}, \Omega_{\Lambda}$ and $H_0$ mentioned in the above section.
For the parameter b  we adopt a convenient prior range as $b \sim -0.001-0.1 .$ In the reference \cite{Guo:2007zk} authors have studied  the observational constraints on a coupling between dark energy and dark matter in which the range for coupling constant is taken as $-0.4<b<0.1$.

The likelihood of the model can be obtained using a relation analogous to 
equation~(\ref{equ:like7})
where we have marginalized over all parameters and the computation has done for all data combinations like model 1 and the values are given in the table ~\ref{table:2}.

\begin{figure}
	\centering
	\includegraphics[scale=0.55]{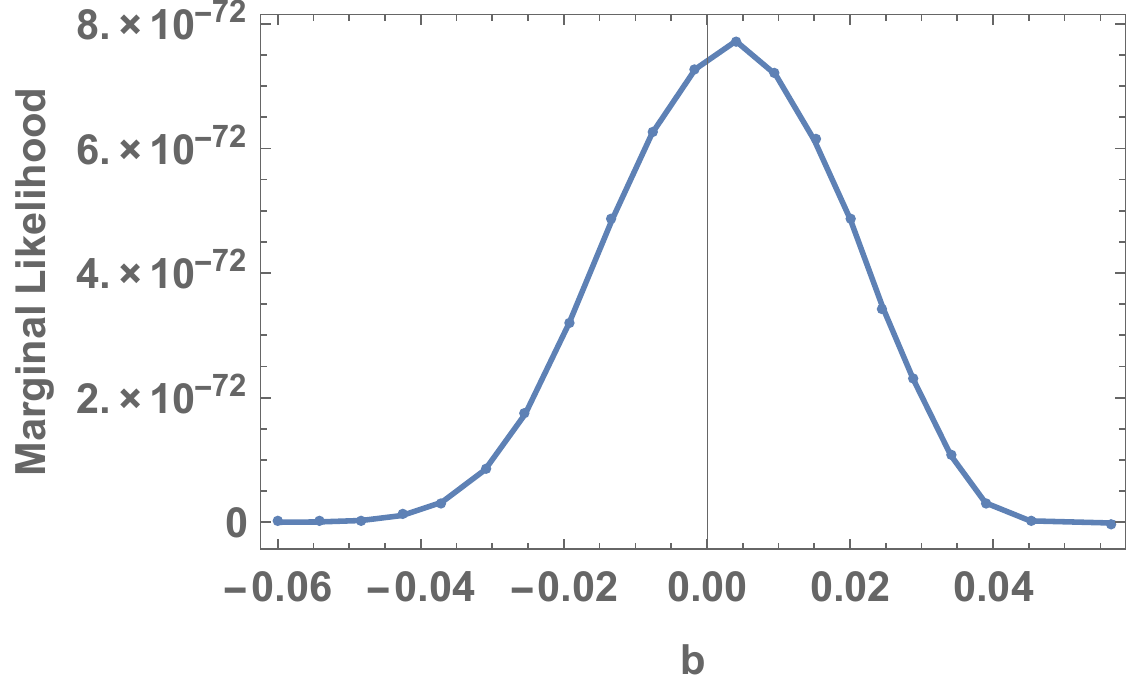}
	\caption{The plot shows the marginal likelihood of the interaction term b in model 2 for SNIa(307)+CMB+BAO data set.}\label{fig:f2}
\end{figure}

\begin{figure}
	\centering
	\includegraphics[scale=0.55]{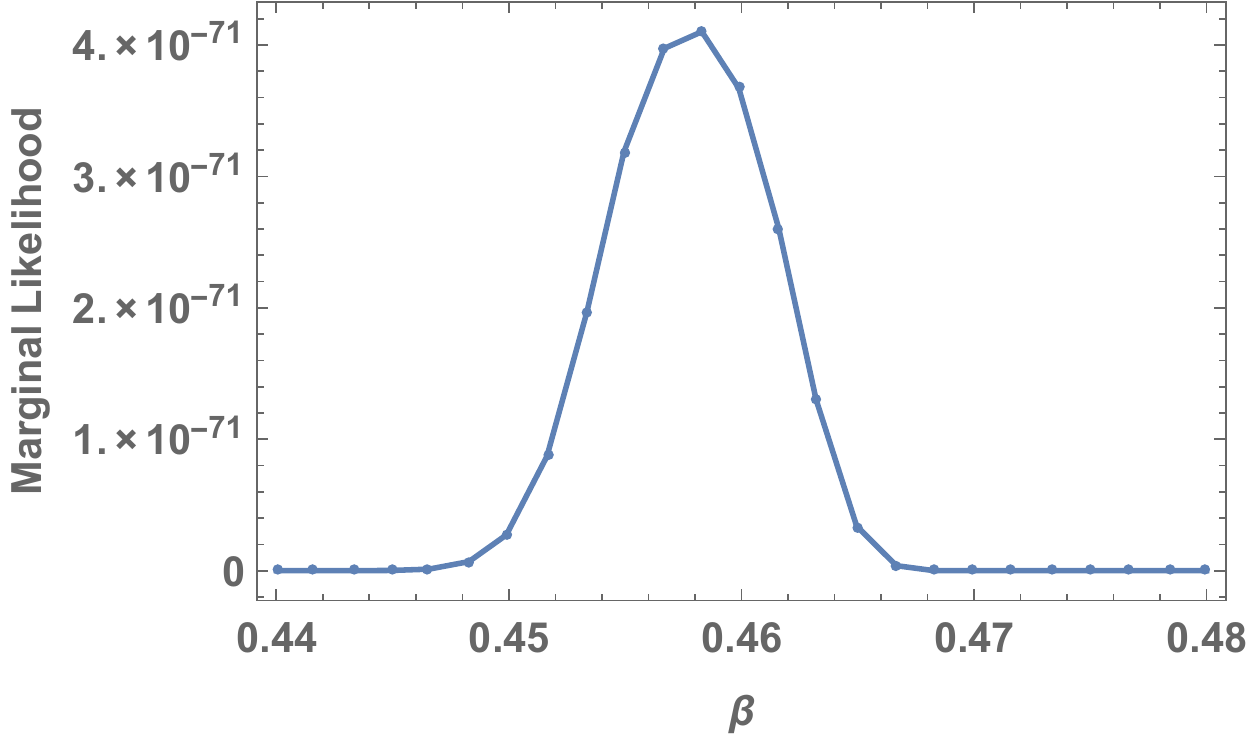}
	\caption{The plot shows the marginal likelihood of the model parameter $\beta$ in model 2  for SNIa(307)+CMB+BAO data sets.}\label{fig:f4}
\end{figure}

The robustness of the calculation can be checked by obtaining the marginal likelihood of the parameters 
$b$ and $\beta.$
The marginal likelihood for the interaction term $b$ can then be obtained by performing the integral over all other parameters as in
relation~(\ref{equ:like3}). 
The resulting function is then plotted against $b$ and is found to have a
Gaussian shape, see  Fig.~(\ref{fig:f2}). The peak value of $b$ in this plot is around 0.0055 and 
is same as the best estimated value of the parameter 
obtained in section 2 by the method of $\chi^2$ minimization.
A similar procedure is adopted for finding the marginal likelihood for the parameter $\beta$ 
and 
the result is shown in Fig.~(\ref{fig:f4}). In this case also the peak of the Gaussian distribution of the parameter, 
$\beta \sim 0.4583$ is same as the one obtained earlier using $\chi^2$ technique.

\begin{table}
	\centering
	\caption{Likelihood and Bayes factor for different data set of $\Lambda$CDM and model 2 in which data1 refers to the combination SNIa(307)+CMB+BAO, data2 refers to SNIa(307)+CMB+BAO+Hubble data, data3  refers to SNIa(580)+CMB+BAO, data4 refers to SNIa(580)+CMB+BAO+Hubble data}\label{table:2}
	\scalebox{1.0}{
		\begin{tabular}{lcccccr}
			\hline
			Data & Likelihood of  & Likelihood of  & Bayes Factor\\ [0.2ex] 
			&   $\Lambda$CDM($\mathcal{L}(M_j)$) &   model 2($\mathcal{L}(M_i)$)		&   $B_{ij}=\frac{\mathcal{L}(M_i)}{\mathcal{L}(M_j)}$&          \\
			\hline
			data1 & 6.8607$\times 10^{-71}$& 5.00575$\times 10^{-73}$ &0.0073 \\ 
			data2 & 2.82795$\times 10^{-80}$ & 2.27908$\times 10^{-82}$& 0.0080 \\
			data3 & 2.51447$\times 10^{-125}$ &1.94276$\times 10^{-127}$ &0.0077\\
			data4 & 2.60201$\times 10^{-135}$ & 2.02187$\times 10^{-137}$&0.0077 \\
			\hline
	\end{tabular}}
\end{table}

The Bayes factor of model 2 corresponding to all the data sets are then calculated by comparing the likelihood of 
model 2 and that of the standard $\Lambda$CDM model, which is the same as used for the case of model 1. 
The results are tabulated in Table~(\ref{table:2}).

According to table 2, 
the Bayes factor of model 2 corresponding to all the data combinations   
are all lie in the range $B_\text{ij}<1.$ According to Jeffreys 
scale this indicate that the evidence for the standard $\Lambda$CDM model is very strong compared to model 2 also with reference to the 
said data combinations.

\section{Conclusion}
The huge discrepancy between the observed value of the cosmological constant and its value predicted by quantum field theory 
is one of the prominent issues in theoretical physics.
A feasible solution to this, as proposed by many, is that the vacuum energy, which is a measure of the cosmological constant, 
can be considered as varying as the universe expands.  According to running dark energy models, the dark energy density 
depends on the Hubble parameter and its derivative\cite{Shapiro:2009dh,PhysRevD.88.063531,doi:10.1142/S0218271815410035}.
This dependency is restricted to the even powers of the Hubble parameter in order to satisfy the general covariance. 
It was noticed that the holographic Ricci
dark energy has the similar density form
as that of the standard running vacuum models \cite{doi:10.1142/S0218271815410035},
except for the difference in the nature of the model parameters and hence it can also be considered as a form of running dark energy,
generically called running holographic Ricci dark energy (rhrde).
In the present work we have considered two types of rhrde and tested their relative significance 
against the standard $\Lambda$CDM model. The first one model 1, describes a two component universe, having rhrde with an additive 
bare cosmological constant in its density and dark matter as cosmic constituents. Model 2, having rhrde without having a bare 
cosmological constant and dark matter as cosmic components, while the interaction between these components is incorporated 
through a non-gravitational phenomenological term,
$Q=3bH\rho_{\text m}.$

We have obtained the likelihood of
model 1, model 2 and also the $\Lambda$CDM model using the different data combinations.
The basic ingredients for the likelihood are the prior ranges for the parameters of the models.
We have used suitable flat priors for different parameters in the two models and are 
$0 \leq \beta \leq 0.5$ for model parameter, 
0.1$ \leq \Omega_{m_0} \leq $0.7 for the present value of the mass density parameter of dark matter,  
$0.2 \leq \Omega_{\Lambda} \leq 1$ for the present value of the mass density parameter of the rhrde and 
$H_0$ in the range 65 - 73km$s^{-1}Mpc^{-1}$ by considering recent Planck observations\cite{2018arXiv180706209P} and the local 
expansion rate estimation\cite{Riess:2016jrr}. 
The ratio of the likelihood of any two models gives the Bayes inference factor.

Our analysis shows that the Bayes factor for model 1 is much less than one for all the data combinations.  As per the 
Jeffreys criterion this shows that the evidence of the standard $\Lambda$CDM is very strong as compared to  model 1.

Model 2, is different from model 1 in two ways; 1. the rhrde doesn't contain the additive bare cosmological constant and 
2. the interaction of dark energy with dark matter is accounted with a phenomenological term, which in fact causes the late acceleration of the universe \cite{George:2018myt}. 
The results of the Bayesian  analysis is summarized in table~\ref{table:2}. It is found that the Bayes factor of model 2 is less than one  and is also found to be less than that for model 1 for all the data combinations. As per the Jeffreys criterion, this range 
of Bayes factor implies that the evidence of the standard $\Lambda$CDM is very strong against model 2 also.

Based on the overall analysis of present work, we can conclude that standard $\Lambda$CDM model is found to have very strong evidence over the running holographic Ricci dark energy models. However at the background level the running vacuum energy models of the universe gives good description of the late universe\cite{Sol__2017,Sola:2016vis,doi:10.1142/S0217751X16300350,Sola:2016ecz}. The author in the reference \cite{Szydlowski:2015bwa} showed that the cosmological model with decaying vacuum is a worse fit than $\Lambda$CDM model. 
At the gross level our analysis also favors this conclusion.
\section*{Acknowledgments}

First of all we are thankful to the Referee for useful comments which helped us in improving this manuscript including the conclusion to a great extend. One of the authors, Paxy George is thankful to DST, Govt. of India, for giving financial support through PURSE
fellowship. The authors wish to thank Moncy V. John for very helpful discussions. The authors also thank Prof. M. Sabir 
for the suggestions on the manuscript. \\[0.15in]

\noindent {\bf Data Availability}
The supernova data underlying this article are available in [SCP "Union2.1" SNIa compilation] at, http://supernova.lbl.gov/Union

\bibliographystyle{aip}
\bibliography{ref6a} 
\end{document}